\documentstyle[11pt]{article}
\begin{document}
{\setlength{\oddsidemargin}{1.2in}
\setlength{\evensidemargin}{1.2in} } \baselineskip 0.55cm
\begin{center}
{ {\bf Quantization of horizon area of Kerr-Newman-de Sitter black hole}}
\end{center}
\begin{center}
 ${\rm Y.\ Kenedy \ Meitei}^{1}$. ${\rm T.\ Ibungochouba\ Singh}^{1}$. ${\rm I.\ Ablu\ Meitei}^{2}$.
\end{center}
\begin{center}
1\ Department of Mathematics, Manipur University, Canchipur, Manipur 795003, India\\
2\ Department of Physics, Modern College, Imphal, Manipur 795005, India\\
E-mail:kendyum123@gmail.com
\end{center}
\date{}

\begin{abstract}
Using the adiabatic invariant action and applying Bohr-Sommerfeld quantization rule and first law of black hole thermodynamics a  study of quantization of entropy and horizon area of Kerr-Newman-de Sitter black hole is carried out. The same entropy spectrum is obtained in two different coordinate systems. It is also observed that the spacing of the entropy spectrum is independent of the black hole parameters and also the corresponding quantum of horizon area is in agreement with the results of Bekenstein.

\end{abstract}

A complete and consistent quantum theory of gravity is still not fully developed though its necessity was recognized as early as 1930s. It is believed that black holes should play an important role in quantum theory of gravity. In the early attempts to quantize black hole, Bekenstei${\rm n}^{[1-4]}$ suggested that in quantum theory the black hole mass spectrum must be discrete and highly degenerate unlike the continuous black hole mass spectrum in general relativity and the black hole surface area spectrum must be discrete and equispaced whose degeneracy corresponds to the black hole entropy associated with area eigenvalue. It was also observed that the horizon area of non-extremal black hole behaves like a classical adiabatic invariant which corresponds to a quantum entity with discrete spectrum. Bekenstei${\rm n}^{[3]}$ found a lower bound on the increase in the black hole surface area in quantum theory which is given by
\begin{eqnarray*}
(\Delta A)_{min}=8\pi \ell^2_{p}
\end{eqnarray*}
where $\ell_{p}=(\sqrt{G/c^3})^{1/2}\hbar^{1/2}$ is the Plank length. Ho${\rm d}^{[5]}$ also found a similar lower bound on the increase of the black hole area caused by the assimilation of charged particle which is given by
\begin{eqnarray*}
(\Delta A)_{min}=4 \ell^2_{p}.
\end{eqnarray*}

Following Bohr's correspondence principle, Ho${\rm d}^{[5]}$ used quasinormal mode frequencies of Schwarzschild black hole as the transition frequency in quantum theory and determined the surface area spectrum which is given by
\begin{eqnarray*}
 A_{n}=(4 \ell^2_{p}\ell n3)n, \,\,\,\,\,\,\,   n=1,2,3....
\end{eqnarray*}
It is characterized by an equidistant area spacing of $4 \ell^2_{p}\ell n3$. Draye${\rm r}^{[6]}$ also found a similar area spacing based on loop quantum gravity.

In a semiclassical study, Padmanabhan and Pate${\rm l}^{ [7]}$ observed that in all static space-times with horizons, the area of the horizon as measured by any observer blocked by that horizon must be quantized. Also, the change in area of the horizon must also be quantized with minimum detectable change of the order of $\ell^2_{p}$.

Based on Hod's ideas, Kunstatte${\rm r}^{[8]}$ considered the quasinormal frequency as a fundamental frequency for a black hole and considering the quantity $I=\int dE/\omega(E) dE$ as an adiabatic invariant where $\omega(E)$ is the vibrational frequency of a system with energy $E$.  The entropy spectrum and hence the area spectrum of a black hole is found to be in agreement with Hod's result.

Maggior${\rm e}^{[9]}$ introduced a new interpretation of the black hole quasinormal modes that the relaxation behavior of  a perturbed black hole governed by the complex quasinormal frequencies $\omega=\omega_R+i\omega_I$ is the same as that of damped harmonic oscillator whose real frequencies are $[\omega^{2}_{R}+\omega^{2}_{I}]^{1/2}$ rather than $\omega_{R}$ used by Hod and the resulting area spectrum is equispaced with $\Delta=8\pi\ell^2_{p}$ which is different from Hod's result but in agreement with the original result of Bekenstein. Vagena${\rm s}^{[10]}$ extended Kunstatter's approach to the case of Kerr black hole using an adiabatic invariant of the form $I=\int(dM-\Omega dJ)/\omega_o$, where $\Omega$ and $J$ are the angular velocity and angular momentum of the black hole respectively and $\omega_o$ is the quasinormal mode frequency of the black hole. Using this adiabatic invariant and considering that in the semiclassical regime, the Kerr black hole must be far away from extremality so that $J<<M^2$, Medve${\rm d}^{[11]}$ showed that Kerr area spectrum is asymptotically identical for the two methods (Hod's and Kunstatter's methods along with modifications suggested by Maggiore and Vagenas) and obtained a universal form for the Schwarzschild and Kerr black hole spectra. Applying quasinormal mode frequencies, the entropy and the area spectra of black holes have been studied in different space-time${\rm s}.^{[12-13]}$ Based on the standard commutation rules and quantization of angular momentum related to the Euclidean Rindler space of black hole Ropotenk${\rm o}^{[14]}$ derived the quantization of black hole horizon area and the area spectrum is applicable to all kinds of black holes. In Rindler space, the Schwarzschild time coordinate transforms to a periodic angle which is conjugate to the z-component of the angular momentum. The resulting spacing of the area spectrum agrees with Bekenstein's result.

In 2011, Majhi and Vagena${\rm s}^{[15]}$ proposed a new approach using an adiabatic invariant quantity of the form $\int p_{i}dq_{i}=\int\int^{H}_{0}\frac{dH'}{\dot{q}_{i}}dq_{i}$  and Bohr-Sommerfield quantization condition where $p_{i}$ is the momentum conjugate to $q_{i}$. In this approach without using the concept of quasinormal modes, the entropy and the horizon area spectrum are derived. These spectra are equispaced and are in agreement with Bekenstein's result.

According to Jiang and Ha${\rm n},^{[16]}$ the adiabatic invariant quantity $\int p_{i}dq_{i}$ is not canonically invariant. Using the covariant action $I=\oint  p_{i}dq_{i}=\oint \frac{dH}{\dot{q}_{i}}dq_{i}$ as an adiabatic invariant quantity, the horizon area of the Schwarzschild black hole is shown to be quantized with equally spaced area spectrum having $\Delta A=8\pi\ell^2_{p}$ in both Schwarzschild and Painlev$\acute{e}$ coordinates. Showing that the time period of outgoing wave is related to the vibrational frequency of the perturbed black hole, Zeng et a${\rm l}^{[17]}$ quantized the horizon area of Schwarzschild and Kerr black holes.  More works have been done to quantize black hole entropy and horizon are${\rm a}.^{[18-23]}$

The KNdS metric in Boyer-Lindquist coordinates can be written a${\rm s}^{[24]}$

\begin{eqnarray}
ds^2&=&-\frac{\Delta-\Delta_\theta a^2\sin^2\theta}{\rho^2\Xi^2}dt^2+\frac{\rho^2}{\Delta}dr^2+\frac{\rho^2}{\Delta_\theta}d\theta^2
-\frac{2a[\Delta_\theta(r^2+a^2)-\Delta]\sin^2\theta}{\rho^2\Xi^2}dtd\phi\cr&&
+\frac{\Delta_\theta(r^2+a^2)^2-\Delta a^2\sin^2\theta}{\rho^2\Xi^2}\sin^2\theta d\phi^2,
\end{eqnarray}
where
\begin{eqnarray}
&&\rho^2=r^2+a^2\cos^2\theta,\,\,\,\,\, \Xi=1+\frac{1}{3}\Lambda a^2,\cr
&&\Delta_\theta=1+\frac{1}{3}\Lambda a^2\cos^2\theta,\cr
&&\Delta=(r^2+a^2)\Big(1-\frac{1}{3}\Lambda r^2\Big)-2Mr+Q^2.
\end{eqnarray}
 This line element represents KNdS solution for $\Lambda>0$, and anti KNdS solution for $\Lambda<0$. M and Q are the mass and the charge of the black hole respectively.
The KNdS solution has singularity when
\begin{equation}
\Delta=r^2+a^2-2Mr+Q^2-\frac{1}{3}\Lambda r^2(r^2+a^2)=0.
\end{equation}
We consider only the case $\Lambda>0$, in which Eq. (3) has four real roots. The case $\Lambda<0$ is ignored, since Eq. (3) gives imaginary roots. The four real roots of Eq. (3), say $r_c$, $r_h$, $r_1$ and  $r_-$ $(r_c>r_h>r_1>0>r_-$) satisfy the relation as
\begin{equation}
(r-r_c)(r-r_h)(r-r_1)(r-r_-)=-\frac{3}{\Lambda}[r^2+a^2-2Mr+Q^2-\frac{1}{3}\Lambda r^2(r^2+a^2)].
\end{equation}
The largest root $r_c$ represents the location of the cosmological horizon, $r_h$ is the location of the black hole event horizon and $r_1$ is the location of the Cauchy horizon. The
smallest negative root, $r_-$ indicates the another cosmological horizon on the other side
of the ring singularity at r = 0 and another infinit${\rm y}.^{[25]}$ We factorize $\Delta$ into the the following for${\rm m}^{[26]}$
\begin{eqnarray}
\Delta=(r-r_h)\Delta'(r_{h})
\end{eqnarray}
where $\frac{d\Delta}{dr}=\Delta'(r_{h})$. The metric (1) has a singularity at the radius of the event horizon. A coordinate system analogus to Painlev${\acute{e}}$ coordinate system will be found out. We see that the new coordinate system is well-behaved at the event horizon and will discuss the dragging coordinate system. Let $\frac{d\phi}{dt}=-\frac{g_{14}}{g_{44}}=\Omega$. The new line element of KNdS is
\begin{eqnarray}
ds^2=\hat{g}_{11}dt^2_k+\frac{\rho^2}{\Delta}dr^2+\frac{\rho^2}{\Delta_\theta}d\theta^2,
\end{eqnarray}
where
\begin{eqnarray}
\hat{g}_{11}=-\frac{\Delta \Delta_\theta\rho^2}{\Xi^2[\Delta_\theta(r^2+a^2)^2-\Delta a^2\sin^2\theta]}.
\end{eqnarray}
The angular velocity at the event of horizon is

\begin{eqnarray}
\Omega_h=\frac{a}{r^2_h+a^2}.
\end{eqnarray}
The surface gravity of the black hole is given by
\begin{eqnarray}
\kappa=\lim_{\hat{g}_{11}\rightarrow 0}\Big(-\frac{1}{2}\sqrt{\frac{-g^{22}}{\hat{g}_{11}}}\frac{d\hat{g}_{11}}{dr}\Big)=\frac{\Delta'(r_h)}{2\Xi(r^2_h+a^2)}.
\end{eqnarray}
In order to study the black hole entropy, the line element of the Euclideanized Kerr-Newman-de Sitter black hole is obtained from Eq. (6) by transforming $t_k\rightarrow -i\tau$ as
\begin{eqnarray}
ds^2=\frac{\Delta\Delta_\theta\rho^2}{\Xi^2[\Delta_\theta(r^2+a^2)^2-\Delta a^2\sin^2\theta]}d\tau^2+\frac{\rho^2}{\Delta}dr^2+\frac{\rho^2}{\Delta_\theta}d\theta^2.
\end{eqnarray}

Now let us consider the adiabatic covariant action for the KNdS black hole. We have seen that the quantum phenomena, such as tunneling radiatio${\rm n},^{[27-32]}$ quantization of black hole entropy and are${\rm a}^{[33-37]}$ are observed at the black hole event horizon. The close contour integral can be used by taking a close path from $q^{out}_{i}$ (outside the horizon) to $q^{in}_{i}$ (inside the horizon), near the event horizon. By adopting this closed integral for the charged black hole, the adiabatic invariant quantity can be expressed as
\begin{eqnarray}
I_{adia}=\oint p_{i}dq_{i}=\int^{q^{out}_{i}}_{q^{in}_{i}}p^{out}_{i}dq_{i}+\int^{q^{in}_{i}}_{q^{out}_{i}}p^{in}_{i}dq_{i},
\end{eqnarray}
where $q_{i}^{out}$'s $(q_{i}^{in})$'s indicate the Euclideanized KNdS spacetime coordinates. $p_{i}^{out}$'s $(p_{i}^{in})$'s represent the canonical momentum of the coordinates $q_{i}^{out}$'s $(q_{i}^{in})$'s. For the quantization of the entropy of the charged black hole, the expression of adiabatic covariant action in the charged black hole will be used. The integral $\int^{q^{out}_{i}}_{q^{in}_{i}}p^{out}_{i}dq_{i}$ can be written as
\begin{eqnarray}
\int^{q^{out}_{i}}_{q^{in}_{i}}p^{out}_{i}dq_{i}=\int^{q^{out}_{i}}_{q^{in}_{i}}\int^{p^{out}_{i}}_{0}dp^{out'}_{i}dq_{i}=
\int^{\tau_{out}}_{\tau_{in}}\int^{H}_{0}dH'd\tau+\int^{q^{out}_{j}}_{q^{in}_{j}}\int^{H}_{0}\frac{dH'}{\dot{q}_{j}},
\end{eqnarray}
where $H$ is the Hamiltonian of the system and ${q_{j}}^{,}s$ represent the space like coordinates. $\int^{\tau_{out}}_{\tau_{in}}\int^{H}_{0}\frac{dH'}{\dot{q}_{j}}$ signifies the classical action $I_c$,  and satisfies $I_c=\int^{\tau_{out}}_{\tau_{in}}Ld\tau$ and $L$ is the Lagrangian function. The effect of the electromagnetic field should be considered when a charge particle tunnels out of the horizon. Then the matter gravity system consists of the black hole and the electromagnetic field outside the black hole. We observe that there is a dragging effect of the coordinate system in the rotating black hole and also in the dragging coordinate system, the matter field in the ergosphere near the horizon should be described. It is also observed that the coordinate $\phi$ does not appear in the dragging metric (6). Therefore, $\phi$ is an ignorable coordinate in the Lagrangian function. For eliminating the two degrees of freedom, the classical action can be written as
\begin{eqnarray}
\int^{q^{out}_{j}}_{q^{in}_{j}}\int^{H}_{0}\frac{dH'}{\dot{q}_{j}}&=&\int^{\tau_{out}}_{\tau_{in}}(L-P_{\phi}\dot{\phi}-P_{A_{\tau}}\dot{A_{\tau}})d\tau,\cr
&=&\int^{r_{out}}_{r_{in}}\int^{p_{r}}_{0} dP'_rdr-\int^{\phi_{out}}_{\phi_{in}}\int^{p_{\phi}}_{0} dP'_\phi d\phi-\int^{A^{out}_{\tau}}_{A^{in}_{\tau}}\int^{P_{A_{\tau}}}_{0} dP'_{P_{A_{\tau}}} dP_{A_{\tau}},\cr
&=&\int^{\tau_{out}}_{\tau_{in}}\Big[\int^{H}_{0}dH'\mid_{(r;\phi,P_\phi;A_{\tau},P_{A_{\tau}})}
-\int^{H}_{0}dH'\mid_{(\phi;r,P_r;A_{\tau},P_{A_{\tau}})}\cr&&-\int^{H}_{0}dH'\mid_{(A_{\tau};r,P_r;\phi,P_\phi)}d\tau\Big],\cr
&=&\int^{\tau_{out}}_{\tau_{in}}\int^{H}_{0}dH' d\tau,
\end{eqnarray}
where $P_{r}, P_{\phi}$ and $P_{A_{\tau}}$ are canonical momenta with respect to $r$, $\phi$ and $A_{\tau}$ and the expression of Hamilton's canonical equations can be expressed as
\begin{eqnarray}
\dot{r}&=&\frac{dH}{dP_r}\mid_{(r;\phi,P_\phi;A_{\tau},P_{A_{\tau}})}, \,\,\,\,\,\,\,\,\,\,\,\,\,\,\,\,dH'\mid_{(r;\phi,P_\phi;A_{\tau},P_{A_{\tau}})}=dM',\cr
\dot{\phi}&=&\frac{dH}{dP_{\phi}}\mid_{(\phi;r,P_r;A_{\tau},P_{A_{\tau}})},\,\,\,\,\,\,\,\,\,\,\,\,\,\,\,\,
dH'\mid_{(\phi;r,P_r;A_{\tau},P_{A_{\tau}})}=\Omega_{h}dJ'\cr
\dot{A_{\tau}}&=&\frac{dH}{dP_{A_{\tau}}}\mid_{(A_{\tau};r,P_r;\phi,P_\phi)},\,\,\,\,\,\,\,\,\,\,\,\,\,\,\,\,
dH'\mid_{(A_{\tau};r,P_r;\phi,P_\phi)}=\Phi_hdQ'.
\end{eqnarray}
Using Eqs. (13) and (14), Eq. (12) can be written as
\begin{eqnarray}
\int p_idq_i=2\int^{r_{out}}_{r_{in}}\int^{(M,J,Q)}_{(0,0,0)}\frac{(dM'-\Omega_{h}dJ'-\Phi_{h}dQ)}{\dot{r}_{out}}dr,
\end{eqnarray}
where $\dot{r}_{in}$ denotes the radial outgoing path when a particle tunnels out, $r_{in}$ and $r_{out}$ are the locations of the outer horizon before and after tunneling. After performing same kind of analysis as above, we get
\begin{eqnarray}
\int^{q^{in}_{i}}_{q^{out}_{i}}p^{in}_{i}=2\int^{r_{in}}_{r_{out}}\int^{(M,J,Q)}_{(0,0,0)}
\frac{(dM'-\Omega_{h}dJ'-\Phi_{h}dQ)}{\dot{r}_{in}}dr,
\end{eqnarray}
where $\dot{r}_{in}$ corresponds to the radial ingoing path during the particle tunneling process. Therefore, the adiabatic covariant action can be written as
\begin{eqnarray}
I_{adia}&=&\oint p_{i}dq_{i}=2\int^{r_{out}}_{r_{in}}\int^{(M,J,Q)}_{(0,0,0)}
\frac{(dM'-\Omega_{h}dJ'-\Phi_{h}dQ)}{\dot{r}_{out}}dr\cr&&+2\int^{r_{in}}_{r_{out}}\int^{(M,J,Q)}_{(0,0,0)}
\frac{(dM'-\Omega_{h}dJ'-\Phi_{h}dQ)}{\dot{r}_{in}}dr.
\end{eqnarray}
Applying the adiabatic covariant action, the area spectrum for the KNdS black hole can be obtained in different coordinate systems.
Applying adiabatic covariant $I_{adia}=\oint p_{i}dq_{i}$ given in Eq. (17), Bohr-Sommerfeld quantization rule and the first law of black hole thermodynamics, the area and the entropy of the KNdS black hole are investigated in dragged coordinate system.
In order to obtain $\dot{r}$, the radial null geodesic is considered. From Eq. (10), the outgoing (ingoing) radial null paths $ds^2=d\theta^2=0$ can be written as
\begin{eqnarray}
\dot{r}\equiv\frac{dr}{d\tau}=\pm i\frac{\Delta\sqrt{\Delta_\theta}}{\Xi\sqrt{[\Delta_\theta(r^2+a^2)^2-\Delta a^2\sin^2\theta]}},
\end{eqnarray}
where positive sign corresponds to the outgoing radial null path and negative sign stands for ingoing null path. That is
\begin{eqnarray}
\dot{r}_{\rm out}=+i\frac{\Delta\sqrt{\Delta_\theta}}{\Xi\sqrt{[\Delta_\theta(r^2+a^2)-\Delta a^2\sin^2\theta]}},\cr
\dot{r}_{\rm in}=-i\frac{\Delta\sqrt{\Delta_\theta}}{\Xi\sqrt{[\Delta_\theta(r^2+a^2)-\Delta a^2\sin^2\theta]}}.
\end{eqnarray}
From Eqs. (17) and (19), the adiabatic covariant action
\begin{eqnarray}
I_{adia}=\oint p_{i}dq_{i}&=&-4i\int^{r_{out}}_{r_{in}}\int^{(M,J,Q)}_{(0,0,0)}\frac{\Xi\sqrt{[\Delta_\theta(r^2+a^2)^2-\Delta a^2\sin^2\theta]}}{\Delta\sqrt{\Delta_\theta}}\cr&&\times(dM'-\Omega_{h}dJ'-\Phi_{h}dQ)dr\cr
&=&4\pi\int^{(M,J,Q)}_{(0,0,0)}\frac{\Xi(r^2_h+a^2)}{\Delta'(r_h)}(dM'-\Omega_{h}dJ'-\Phi_{h}dQ)\cr
&=&\int^{(M,J,Q)}_{(0,0,0)}\frac{(dM'-\Omega_{h}dJ'-\Phi_{h}dQ)}{T}.
\end{eqnarray}
From the first law of black hole thermodynamics, we get
\begin{eqnarray}
dM=TdS+\Omega_{h}dJ+\Phi_{h}dQ,
\end{eqnarray}
the adiabatic covariant action in terms of entropy is given by
\begin{eqnarray}
I_{adia}=\oint p_{i}dq_{i}=\hbar S.
\end{eqnarray}
Implementing the Bohr-Sommerfeld quantization rule
\begin{eqnarray}
\oint p_{i}dq_{i}=nh=2\pi n \hbar,  \,\,\,\,\,\,\,\,\, n=1, 2, 3 ,...
\end{eqnarray}
 the entropy spectrum of the KNdS black hole is given by
\begin{eqnarray}
S=2\pi n  \,\,\,\,\,\,\,\,\, n=1, 2, 3 ...
\end{eqnarray}
and the spacing of the entropy spectrum
\begin{eqnarray}
\triangle S=S_{n+1}-S_{n}=2\pi.
\end{eqnarray}
It implies that the entropy spectrum is discrete for the KNdS black hole. The relation between entropy and horizon area is given by
\begin{eqnarray}
S=\frac{A}{4\ell^2_p},
\end{eqnarray}
and the spacing of the horizon area is given by
\begin{eqnarray}
\Delta A=8\pi\ell^2_p,
\end{eqnarray}
where $\ell_{p}=(\sqrt{G/c^3})^{1/2}\hbar^{1/2}$ stands for the Plank length. Therefore, the entropy spectrum is equally spaced and independent of the parameters of KNdS black hole and is in agreement with Bekenstein's results.

 Using adiabatic covariant action $I_{adia}=\oint p_{i}dq_{i}$ given in Eq. (23), Bohr-Sommerfeld quantization rule and the first law of black hole thermodynamics, we will investigate the entropy and horizon area of the KNdS black hole in the dragged-Painlev$\acute{e}$ coordinate system. Introducing Painlev$\acute{e}$ type coordinate transformatio${\rm n}^{[38]}$
\begin{eqnarray}
dt_{k}=dT+F(r,\theta)dr+G(r,\theta)d\theta,
\end{eqnarray}
where
\begin{eqnarray}
 F^2(r,\theta) =\frac{\Xi^2[\Delta_\theta(r^2+a^2)^2-\Delta a^2\sin^2\theta]}{\Delta^{2}\Delta_\theta\rho^2}
 \end{eqnarray}
 and
 \begin{eqnarray}
 G(r,\theta)=\int\frac{\partial F(r,\theta)}{\partial\theta}dr+D(\theta).
 \end{eqnarray}
 We obtain the KNdS black hole metric in dragged-Painlev$\acute{e}$ coordinate
\begin{eqnarray}
ds^2&&=-\frac{\Delta\Delta_\theta \rho^2}{\Xi^2 [\Delta_\theta(r^2+a^2)^2-\Delta a^2\sin^2\theta]}dT^2+\frac{2}{\Xi}\sqrt{\frac{\rho^2\Delta_\theta(\rho^2-\Delta)}{[\Delta_\theta(r^2+a^2)^2-\Delta a^2\sin^2\theta]}}dTdr\cr&&-\frac{2G(r,\theta)\Delta\Delta_\theta \rho^2}{\Xi^2 [\Delta_\theta(r^2+a^2)^2-\Delta a^2\sin^2\theta]}dT d\theta+\frac{\rho^2}{\Delta_\theta}[1-\frac{\Delta \Delta^2_\theta G^2(r,\theta)}{[\Delta_\theta(r^2+a^2)^2-\Delta a^2\sin^2\theta]}]d\theta^2\cr&&+dr^2
+\frac{2G(r,\theta)}{\Xi}\sqrt{\frac{(\rho^2-\Delta)\rho^2\Delta_\theta}{[\Delta_\theta(r^2+a^2)^2-\Delta a^2\sin^2\theta]}}dr d\theta.
\end{eqnarray}
 The components of metric (31) are well behaved not diverging at the horizon and is a flat Euclidean space in radial to constant time slice and also $\partial_T$ is a time like Killing vector field, which keeps the spacetime stationary. The line element (31) should satisfy Landau's condition of coordinate clock synchronization.
 In the dragged Painlev$\acute{e}$ coordinate system, the Euclideanized metric is obtained by a transformation $T\rightarrow -i\tau'$ in the metric (31). Therefore, the corresponding radial null geodesic metric is given by
\begin{eqnarray}
\dot{r}=i\frac{\sqrt{\Delta_\theta}}{\Xi\sqrt{\Delta_\theta(r^2+a^2)^2-\Delta a^2\sin^2\theta]}}[\pm\rho^2-\sqrt{\rho^2(\rho^2-\Delta)}],
\end{eqnarray}
where the positive (negative) sign stands for the outgoing (ingoing) radial null paths, namely
\begin{eqnarray}
\dot{r}_{out}&=&i\frac{\sqrt{\Delta_\theta}}{\Xi\sqrt{\Delta_\theta(r^2+a^2)^2-\Delta a^2\sin^2\theta]}}[\rho^2-\sqrt{\rho^2(\rho^2-\Delta)}]\cr
\dot{r}_{in}&=&-i\frac{\sqrt{\Delta_\theta}}{\Xi\sqrt{\Delta_\theta(r^2+a^2)^2-\Delta a^2\sin^2\theta]}}[\rho^2+\sqrt{\rho^2(\rho^2-\Delta)}].
\end{eqnarray}
Inserting Eq. (33) into Eq. (17), we get
\begin{eqnarray}
I_{adia}&=&\oint p_{i}dq_{i}\cr
&=&-2i\int^{r_{out}}_{r_{in}}\int^{(M,J,Q)}_{(0,0,0)}\frac{\Xi\sqrt{[\Delta_\theta(r^2+a^2)^2-\Delta a^2\sin^2\theta]}(dM'-\Omega_{h}dJ'-\Phi_{h}dQ)}{\sqrt{\Delta_\theta}[\rho^2-\sqrt{\rho^2(\rho^2-\Delta)}]}dr\cr&&
+2i\int^{r_{out}}_{r_{in}}\int^{(M,J,Q)}_{(0,0,0)}\frac{\Xi\sqrt{[\Delta_\theta(r^2+a^2)^2-\Delta a^2\sin^2\theta]}(dM'-\Omega_{h}dJ'-\Phi_{h}dQ)}{\sqrt{\Delta_\theta}[\rho^2+\sqrt{\rho^2(\rho^2-\Delta)}]}dr.\nonumber\\
\end{eqnarray}
Note that the integral of second term in Eq. (34) has no contribution in the dragged Painlev$\acute{e}$ coordinat${\rm e}.^{[22]}$ Therefore the adiabatic covariant action can be written as
\begin{eqnarray}
I_{adia}&=&\oint p_{i}dq_{i}\cr
&=&-2i\int^{r_{out}}_{r_{in}}\int^{(M,J,Q)}_{(0,0,0)}\frac{\Xi\sqrt{[\Delta_\theta(r^2+a^2)^2-\Delta a^2\sin^2\theta]}(dM'-\Omega_{h}dJ'-\Phi_{h}dQ)}{\sqrt{\Delta_\theta}[\rho^2-\sqrt{\rho^2(\rho^2-\Delta)}]}dr\cr
&=&4\pi\int^{(M,J,Q)}_{(0,0,0)}\frac{\Xi(r^2_h+a^2)(dM'-\Omega_{h}dJ'-\Phi_{h}dQ)}{\Delta'(r_h)}.
\end{eqnarray}
 Using the first law of black hole thermodynamics, $dM=TdS+\Omega_h dJ+\Phi_h dQ$ and Bohr-Sommerfeld quantization rule, $\oint p_{i}dq_{i}=nh$, the quantization of the entropy is obtained as
 \begin{eqnarray}
 S=2\pi n,
\end{eqnarray}
where $n=1,2,3,...$ and the spacing of the entropy spectrum is
\begin{eqnarray}
\Delta S=2\pi.
\end{eqnarray}
Utilizing the relation between entropy and horizon area
\begin{eqnarray}
\Delta S=\frac{\Delta A}{4\ell^2_p}.
\end{eqnarray}
The spacing of the area spectrum is given by
\begin{eqnarray}
\Delta A=8\pi\ell^2_p.
\end{eqnarray}

Hence, we have recovered Bekenstein's original result in the two different coordinates, namely, the dragged spherical coordinate and the dragged Painlev$\acute{e}$  coordinate. The results show that the black hole area spectrum is uniformly spaced and independent of the parameters of the black hole e.g. mass $M$, angular momentum $J$ and charge $Q$ in agreement with the earlier results based on different technique${\rm s}.^{[9-11, 17, 39]}$ The universal form for the spectra of Schwarzschild and Kerr black holes observed by Vagena${\rm s}^{[11]}$ has been extended to Kerr-Newman-de Sitter black hole.

In this paper, we study the quantization of horizon area and entropy of KNdS black hole in different coordinates using the modified adiabatic covariant action. In order to study spectroscopy of black hole, the expression of the adiabatic invariant quantity in the dragged coordinate system is presented. Then, via modified adiabatic covariant action, implementing the Bohr-Sommerfeld quantization rule and the first law of black hole thermodynamics, the entropy spectra and the area of the KNdS black hole in different coordinates are obtained near the horizon.
These results imply that the entropy spectrum and the area spectrum are equally spaced and do not depend on the black hole parameters. Our work is consistent with the already obtained result in the literature by Maggiore with the quasinormal modes and it also confirmed the initial result obtained by Bekenstein. Using the dragged spherical coordinate and dragged Painlev$\acute{e}$ coordinate, via the modified adiabatic covariant action,  we obtain the same value of area and entropy spectra of black hole, which is a physically required result since the area spectrum should be invariant under the coordinates transformations. This technique gives a suitable method to quantize the horizon area for various black holes, specially for a complicated background spacetime like KNdS black hole.

{\bf Acknowledgements} : YKM acknowledges CSIR for providing financial support.

\end{document}